\documentclass[12pt]{iopart}
\usepackage{iopams}
\usepackage{setstack}
\usepackage{amsfonts}
\begin{document}

\title[]{Quantum logic under semi-classical limit: information loss}

\author{Maksym Teslyk, Olena Teslyk}

\address{Faculty of Physics, Taras Shevchenko National University of Kyiv, Kyiv, Ukraine}
\ead{machur@ukr.net}
\vspace{12pt}

\begin{abstract}
	We consider quantum computation efficiency from a new perspective. The efficiency is reduced to its classical counterpart by imposing the semi-classical limit. We show that this reduction is caused by the fact that any elementary quantum logic operation (gate) suffers information loss during transition to its classical analogue. Amount of the information lost is estimated for any gate from the complete set. The largest loss is obtained for non-commuting gates that allows to consider them as quantum computational speed-up resource. Our method allows to quantify advantages of quantum computation as compared to the classical one by direct analysis of the basic logic involved. The obtained results are illustrated by application to quantum discrete Fourier transform and Grover search algorithms.
\end{abstract}

\vspace{2pc}


\maketitle
%
%

\section{Introduction}
\label{sec:intro}
The construction of a quantum computer is an important open problem
in modern physics. The interest in this endeavour is mainly due to
high efficiency of quantum algorithms, such as (but not only) the
Grover search and Shor's factoring, and the fact that their
classical analogues are much less efficient. But why are quantum
calculations much more efficient than the classical analogues? The
common answer to this question is: the speed-up is based on the
quantum parallelism and, probably, on entanglement. However this is
only the qualitative explanation, and it is reasonable to try to
explain the gap in efficiency from the basic principles of quantum
and classical computation. Every computation, either it is quantum
or classical, can be decomposed into a set of elementary operations.
We impose the semiclassical limit and study how the complete set of
quantum gates is reduced to the classical counterpart. We base our
analysis on the formal rules of the quantum and classical logics,
which were first formulated by G. Birkhoff and J. von Neumann in
their seminal paper \cite{ql_Neumann}.

To date much progress has been done in the area; for the reviews see \cite{ql_rev,ql_rev_2}. Some possible quantum computational
structures are presented in \cite{ql_comp}. \cite{ql_no-commut} is devoted to investigations in the algebraic
structure of logic within the framework of non-commutative geometry.
In \cite{ql_m-alg} the so-called measurement algebras, the
formalism of which is weaker than that of Hilbert spaces, were
explored. Description of the orthomodular lattices via the
Sasaki projection is presented in \cite{ql_orthoposet}.
\cite{ql_context,ql_contextual} are devoted to the
analysis of contexts, i.e. the maximal sets of commuting logic
statements. Different approaches in formal representation and
formalism initiation for quantum logic (QL) have been explored.
Investigations in categorical QL are presented in \cite{ql_categ,ql_categ_sem}. On the measurement-based QL and
computation, which are strongly connected to projective operators
logic representation \cite{ql_Neumann-2}, we refer to \cite{ql_rev_meas,ql_meas}, where the first one is devoted to the
``reversible measurement'' \--- a hypothetical operation allowing to
``look inside'' the quantum computation, and the second one
describes measurement-based computation on graph states. Theory of finite automata based on QL
one may find in \cite{ql_l-val}. QL may be interpreted as a language of ``pragmatically'' decidable
assertive formulas, thus formalizing statements about physical
quantum systems \cite{ql_pragmatic}. \cite{ql_and-then} is
devoted to QL representation based on the alternative set of logic
operations. In \cite{ql_lambda} author expands the
$\lambda$-calculus on quantum computation. For some extensions of QL
we refer to \cite{ql_temp,ql_en_conn,ql_mod,ql_mod_2}. Computational
complexity in quantum and classical logic (CL) calculus are explored
in \cite{qc_k_i_def} or others, such as \cite{qc_k_intro,qc_k_alg,qc_k_appl,it_alg_e}. Attempts in
bridging semantic space and QL are considered in \cite{ql_sem}. Quantum language investigation was made in
\cite{ql_q-langs}.

Thus we conclude that investigations in logic (especially quantum) are
rather actual and are tightly interconnected spanning different
areas of research.

It is legitimate to conclude that the QL is much more efficient than
the CL. Such a conclusion is confirmed by the fact that the
algebraic structure of QL is constructed with the help of weaker
conditions than that of CL, thus allowing a wider class of
operations to be processed. Here we interpret efficiency in the
computational complexity sense, i.e. as the number of elementary
logical operations necessary to execute some algorithm.

This paper aims to ascertain reasons for the QL being much more
efficient than the CL in terms of a quantitative, rather than the
qualitative (which is mentioned above) explanation of its
superiority. The main motivation of our research is that the
existing approaches and techniques can not completely explain the
efficiency gap. Till now there is no complete theory of the
classical and quantum complexity classes and of interrelations
between them. We hope that the approach presented in the manuscript
will be helpful in this challenging problem.

In our research we study how the set of elementary QL operations
reduce to the classical counterpart under taking the semi-classical
limit $\hbar\to0$ (for brevity in the following  we use the term
`dequantization' to denote the semiclassical limit). We apply the
projective operator representation of QL, see \cite{ql_Neumann-2}. One may argue that the projector sets are
not as wide-spread in quantum information research as the other sets
of gates (such as one-qubit gates and Toffoli gate or Controlled-NOT
for example). We use the projectors since we are going to
investigate the formal rules of both quantum and classical logics.
It is much more easier to do that on the similar formal sets of
logic gates such as conjunction, implication and negation than to
try to find a classical analog for Walsh-Hadamard gate for example.
In addition, as it follows from the algorithm complexity theory, the
choice of language is not essential.

After that we estimate the amount of information loss during
dequantization process, thus shedding light on the loss of logic
efficiency and on the efficiency gap problem itself. To quantify our
approach, we use the von Neumann and Shannon entropies for the
quantum and dequantized logic gates respectively.

Estimation of the information difference in QL and CL can be made by
means of the well-known Kolmogorov complexity or quantum
complexities \cite{qc_k_i_def,qc_k_intro,qc_k_alg}. Some common
properties of them and their possible applications were studied in
\cite{qc_k_appl}. Alternatively, algorithmic entropies can
be applied \cite{it_alg_e}.

Our method, having much in common with the Kolmogorov complexity, is
however different. We estimate the information loss of every
elementary logic operation during \emph{reduction} of QL to CL, and
then generalize to an arbitrary calculation. Such an approach allows
to estimate the contribution to the quantum (classical) calculation
of any subspace (domain) of Hilbert (phase) spaces correspondingly.

\cite{ql_deq_improper,ql_deq_conv} extend QL proposed in
\cite{ql_Neumann}. We build upon these studies by providing
dequantization of the complete set of logic operations. To do this,
we use path integral formalism together with the von Neumann and
Shannon entropy definitions. It allows to estimate the information
loss of any quantum algorithm in the semiclassical limit. Compared
to \cite{ql_deq_path_int}, we go further and formalize the
approach for any logic gate.

The interrelation between abelian QL subalgebras and CL algebra has been
explored in \cite{ql_deq_sheaf}. In \cite{ql_deq_lift} some aspects of dequantization of measurement
and of entanglement, which is noted as lifting, were considered with
the help of logic entropy. Instead, we consider dequantization of
any QL statement using the von Neumann and Shannon entropy
definitions. We demonstrate that the non-commuting propositions play
significant role in QL efficiency. Compared to the results of Gottesman-Knill theorem, we exactly demonstrate the significant efficiency contribution of non-commuting statements, which may be outside the Clifford group, while making the transition from QL to CL.

The method we propose estimates the amount of information loss (IL)
for every elementary logical operation after its processing through
the semiclassical limit, but not in the register itself. Efficiency
in its common interpretation is invariant under dequantization,
since the amount of elementary logical operations does not change in
the limit $\hbar\to0$.

In this paper we develop the general scheme of IL estimation for any
QL proposition.

Finally, we exemplify the obtained results with the dequantization
and IL estimation of quantum discrete fast Fourier transform
(FFT$_\mathrm{Q}$) and Grover search (Gr$_\mathrm{Q}$) algorithms.
It was demonstrated \cite{id_repr,id_cplx} that FFT$_\mathrm{Q}$
transforms into the Legendre transform under the dequantization. It
indicates that the dequantized algorithm might change the task it
solves.

We stress that our method does not allow to compare the quantum and
classical algorithms directly: under dequantization the initial task that the algorithm has been designed for may change. The approach calculates amount of
information being lost with any quantum logic proposition under dequantization only, i.e. it is useful in analysis of how much computational efficiency is being lost under the semi-classical transition. Application for complexity classification requires reverse engineering of the approach: one should be able to estimate the gain in efficiency for any classical algorithm while transiting it to the quantum one; such a task is highly non-trivial to date, see Section~\ref{sec:dis} for details.

In Sections~\ref{sec:cl} and \ref{sec:ql} we introduce the CL and QL
formalisms correspondingly; for details see \cite{ql_Neumann,ql_Neumann-2}. Dequantization of QL operations
one can find in Section~\ref{sec:deq}. Estimation of the information
loss during the transition from QL to CL is presented in Section~\ref{sec:eff}.
We formulate and prove theorem, which is necessary for the
application of the technique to any quantum algorithm, in Section~\ref{sec:thm}.
Examples of how the scheme works on FFT$_\mathrm{Q}$
and Gr$_\mathrm{Q}$ are given in Section~\ref{sec:ex}. Discussion of
the obtained results, their relation to other approaches and open
questions one may find in Section~\ref{sec:dis}.

\section{Classical Logic}
\label{sec:cl}
Let $\Gamma_S$ be the phase space describing a physical system $S$
in some state $\lambda$. We assume that this state corresponds to
some domain in $\Gamma_S$ and it is characterized by a
characteristic function $\chi_{\lambda}$ which is defined on
$\Gamma_S$. The statement ``$S$ possesses physical property
$\lambda$'' or ``$S$ is in the state $\lambda$'' will be true or
false for those domains in $\Gamma_S$ where $\chi_{\lambda}=1$ or
$\chi_{\lambda}=0$ respectively.

Such characteristic functions may be used to define formal rules and
elementary operations of CL on $\Gamma_S$. For example, they can
describe conjunction, implication and negation in terms of the phase
space subsets \cite{ql_Neumann}.

Conjunction $\wedge$ is defined as
\begin{equation}\label{eq:c-conj}
\chi_\wedge=\chi_{\lambda}\wedge\chi_{\mu} =
\chi_{\lambda}\chi_{\mu}
\end{equation}
and describes the intersection subset.

Implication $\leq$ is defined as
\begin{equation}\label{eq:CL_impl}
\chi_{\lambda}\leq\chi_{\mu}\colon\chi_{\lambda}\wedge\chi_{\mu}=\chi_{\lambda}
\end{equation}
and corresponds to rules of the subset inclusion; this operation
initiates statement ordering.

Negation $\neg$
\begin{equation}\label{eq:CL_neg}
\chi_{\neg\lambda} = 1-\chi_{\lambda}
\end{equation}
is equivalent to transition to the complementing subset.

Also the operation of disjunction $\vee$ may be introduced. However,
as $\vee$ can be expressed in terms of preliminary operations
\begin{equation}\label{eq:CL_disj}
\chi_\vee=\chi_{\lambda} + \chi_{\mu} - \chi_{\lambda}\chi_{\mu},
\end{equation}
it is not important for us in the following.

\section{Quantum Logic}
\label{sec:ql}
Let ${\rm H}_S$ be the Hilbert space of a physical system $S$. Let $S$ be
in a state $|\zeta\rangle$. Then for any statement about some
property $\lambda$ of $S$ there exists a projective operator
$\mathrm{P}_{\lambda}$ projecting its state onto the corresponding
subspace of ${\rm H}_S$. In other words, the statement ``$S$ possesses
physical property $\lambda$'' will be true if
$\mathrm{P}_{\lambda}|\zeta\rangle\neq0$, and false if
$\mathrm{P}_{\lambda}|\zeta\rangle=0$.

The projective operators on ${\rm H}_S$ have much in common with the
classical characteristic functions on $\Gamma_S$. However, there are
some significant differences: $\mathrm{P}_{\lambda}$ defines some
subspace in ${\rm H}_S$, while $\chi_{\lambda}$ defines some domain in
$\Gamma_S$; two projective operators do not commute in general, but
any two characteristic functions do.

To start with, let us define quantum logical operations for
commuting projectors.

Conjunction $\wedge$ is defined as
\begin{equation}\label{eq:QL_c-conj}
\mathrm{P}_{\wedge}|\zeta\rangle = \left(\mathrm{P}_{\lambda}\wedge
\mathrm{P}_{\mu}\right)|\zeta\rangle =
\mathrm{P}_{\lambda}\mathrm{P}_{\mu}|\zeta\rangle =
\mathrm{P}_{\mu}\mathrm{P}_{\lambda}|\zeta\rangle.
\end{equation}
It describes the intersection of subspaces of \emph{commuting}
operators.

Implication $\leq$ is defined as
\begin{equation}\label{eq:QL_impl}
\mathrm{P}_{\lambda}\leq \mathrm{P}_{\mu}\colon
\left(\mathrm{P}_{\lambda}\wedge
\mathrm{P}_{\mu}\right)|\zeta\rangle =
\mathrm{P}_{\lambda}|\zeta\rangle\quad\forall\,|\zeta\rangle.
\end{equation}
It corresponds to the subspace inclusion. It initiates the statement
ordering similarly to its classical analog. The same definition will
also hold true for non-commuting projectors.

Negation $\neg$ (complementation) is defined as
\begin{equation}\label{eq:QL_neg}
\mathrm{P}_{\neg\lambda}|\zeta\rangle =
\left(\mathrm{I}-\mathrm{P}_{\lambda}\right)|\zeta\rangle,
\end{equation}
where $\mathrm{I}$ is the unit operator. This operation is
equivalent to transition to the orthogonal subspace.

We define conjunction for non-commuting operators as (see \cite{ql_context}, table IV)
\begin{equation}\label{eq:QL_nc-conj}
\mathrm{P}_\wedge|\zeta\rangle = \left(\mathrm{P}_{\lambda}\wedge
\mathrm{P}_{\mu}\right) |\zeta\rangle =
\lim_{n\to\infty}\left(\mathrm{P}_{\lambda}\mathrm{P}_{\mu}\right)^n|\zeta\rangle.
\end{equation}
Such a definition is necessary since conjunction leaves the
statement belonging to both subspaces only, which are determined by
$\mathrm{P}_\lambda$ and $\mathrm{P}_\mu$ (see \cite{ql_context} for details). One can easily verify that such a
definition ensures that conjunction is a projective operator, i.e.
$\mathrm{P}_\wedge^2=\mathrm{P}_\wedge$. Obviously, if
$\mathrm{P}_{\lambda}\mathrm{P}_{\mu}=\mathrm{P}_{\mu}\mathrm{P}_{\lambda}$,
then \eref{eq:QL_nc-conj} transforms into \eref{eq:QL_c-conj}.

Similarly to the CL case, disjunction may be expressed in terms of
the previously defined operations
\begin{equation}\label{eq:QL_disj}
\mathrm{P}_\vee|\zeta\rangle =
\left(\mathrm{P}_{\lambda}+\mathrm{P}_{\mu}-\mathrm{P}_{\lambda}\wedge
\mathrm{P}_{\mu}\right)|\zeta\rangle
\end{equation}
and thus is not needed in the following.

\section{Quantum Logic Dequantization}
\label{sec:deq}
Let $|\zeta\rangle$ be any state in the Hilbert space ${\rm H}_S$ of the
physical system $S$. Projective operator $\mathrm{P}_{\lambda}$
projects the state onto some subspace in ${\rm H}_S$. Within the path
integral formalism it can be written as
\begin{equation}\label{eq:QL_proj}
\mathrm{P}_{\lambda}|\zeta\rangle =
|\lambda\rangle\langle\lambda|\zeta\rangle =
\int\mathcal{D}xe^{iS_{\lambda}[x]/\hbar}
\int\mathcal{D}ye^{iS_{\lambda\to\zeta}[y]/\hbar},
\end{equation}
where integration is made over all phase space trajectories
possible. Here $S_{\lambda}[x]$ is the action describing transition
to the state $|\lambda\rangle$ (that is underlined with the
subscript $_{\lambda}$) along some fixed trajectory $ x $ in phase space with $ x = \{x_1,x_2,x_3,p_{x_1},p_{x_2},p_{x_3}\} $. Action
$S_{\lambda\to\zeta}[y]$ describes transition amplitude
$\langle\lambda|\zeta\rangle$ (that is underlined with a
subscript $_{\lambda\to\zeta}$) along some fixed phase trajectory
$ y $ with $ y = \{y_1,y_2,y_3,p_{y_1},p_{y_2},p_{y_3}\} $.

Result of projection $ \mathrm{P}_{\lambda}|\zeta\rangle $ may be interpreted as state $ |\lambda\rangle $ multiplied by the corresponding transition amplitude $ \langle\lambda|\zeta\rangle $ . In terms of path integration over the phase space this process may be represented as consisting of two stages: integration over the phase space resulting in the state $ |\lambda\rangle $ (stage 1, the 1st path integral) multiplied by integration over the phase space resulting in $ \langle\lambda|\zeta\rangle $ (stage 2, the 2nd path integral).

Such a representation of projective operator has much in common with
symbol of operator. It interconnects $\mathrm{P}_\lambda$ (operator)
defined in Hilbert space to the action (symbol of operator) defined
in phase space.

Taking the limit $\hbar\to0$ results in classical action. Path
integrals extinct when taking the limit because of fast oscillating
exponents, and only trajectories for which action has the extremum
survive. It gives
\begin{eqnarray*}
\lim_{\hbar\to0}&&\frac{\hbar}{i}\ln\int\mathcal{D}x
e^{iS_{\lambda}[x]/\hbar}\int\mathcal{D}y
e^{iS_{\lambda\to\zeta}[y]/\hbar}\nonumber\\
    &=&\cases{
    			S_{\lambda}[x] + S_{\lambda\to\zeta}[y]& for $ \delta S_{\lambda}[x]=\delta S_{\lambda\to\zeta}[y]=0 $\\
    			0& for $ \delta S_{\lambda}[x]\neq0 $ or $ \delta S_{\lambda\to\zeta}[y]\neq0 $\\
    		},
\end{eqnarray*}
where $\delta$ is variation. So one obtains that
\[
\lim_{\hbar\to0}\frac{\hbar}{i}\ln \mathrm{P}_{\lambda}|\zeta\rangle
= \left(S_{\lambda}[x]+S_{\lambda\to\zeta}[y]\right)\chi_{\lambda},
\]
where
\begin{eqnarray*}
\chi_{\lambda}=\cases{
						1&for $ \delta S_{\lambda}[x]=\delta S_{\lambda\to\zeta}[y]=0 $\\
						0&for $ \delta S_{\lambda}[x]\neq0 $ or $ \delta S_{\lambda\to\zeta}[y]\neq0 $\\
					},
\end{eqnarray*}
or in the compact form
\begin{equation}\label{eq:QL_proj_deq}
P_{\lambda}|\zeta\rangle
\overset{\hbar\to0}{\dashrightarrow}\chi_{\lambda}.
\end{equation}
Expression \eref{eq:QL_proj_deq} defines the transition from the
projective operator $\mathrm{P}_\lambda$ to some characteristic
function $\chi_\lambda$. The notation $\chi_{\lambda}$ is used
because $|\zeta\rangle$ is any vector from ${\rm H}_S$ and so there is no
need in the subscript $_\zeta$. This function  defines the classical
action that describes transition of $S$ from the state with some
physical property $\lambda$ to the state with the property $\zeta$.
As one can see, $\chi_\lambda$ vanishes only for those regions in
the phase space where
$\delta\left(S_{\lambda}[x]+S_{\lambda\to\zeta}[y]\right)\neq0$.

Expression \eref{eq:QL_proj_deq} encodes the transition of the
system's description from the quantum mechanical to the classical
one. At the beginning one has the Hilbert space with projectors and
wavefunctions, see \eref{eq:QL_proj}, and at the end one obtains the
phase space with some classical trajectories fixed by the extremum
of the action. The transition (we call it qeduantization for
brevity) is similar to a well-known semiclassical approximation,
when the wavefunction is being expanded into series by $\hbar$ up to
the zeroth order.

At first we consider QL operations for commuting projectors.

Conjunction of two commuting operators
\[
\left(\mathrm{P}_{\lambda}\wedge
\mathrm{P}_{\mu}\right)|\zeta\rangle =
|\lambda\rangle\langle\lambda|\mu\rangle\langle\mu|\zeta\rangle
\]
after taking the limit $\hbar\to0$ \eref{eq:QL_proj_deq} transforms as
\begin{equation}\label{eq:QL_c-conj_deq}
\left(\mathrm{P}_{\lambda}\wedge
\mathrm{P}_{\mu}\right)|\zeta\rangle\overset{\hbar\to0}{\dashrightarrow}
\chi_{\lambda}\chi_{\mu},
\end{equation}
that corresponds to the classical conjunction \eref{eq:c-conj}.

Negation \eref{eq:QL_neg} can be written as
\[
\mathrm{P}_{\neg\lambda}|\zeta\rangle = \left(\mathrm{I} -
\mathrm{P}_{\lambda}\right)|\zeta\rangle =
\left(\int_{-\infty}^{+\infty} |\mu\rangle\langle\mu|\mathrm{d}\mu -
|\lambda\rangle\langle\lambda|\right)|\zeta\rangle,
\]
thus giving the equivalent classical expression, see \eref{eq:CL_neg},
\begin{equation}\label{eq:QL_neg_deq}
\mathrm{P}_{\neg\lambda}|\zeta\rangle\overset{\hbar\to0}{\dashrightarrow}
1-\chi_{\lambda}.
\end{equation}

Now we consider conjunction of non-commuting operators. This
case is more complicated because of appearance of the commutator in
expressions. Dequantization will consist of two steps: at first any
power of product of two non-commuting projective operators will be
considered, and only then their conjunction will be dequantized.

Let $P_{\lambda},P_{\mu}$ be two non-commuting projective operators
such that
\begin{equation}\label{eq:QL_proj_commutator}
\mathrm{P}_{\lambda}\mathrm{P}_{\mu}-\mathrm{P}_{\mu}\mathrm{P}_{\lambda}=i\hbar\mathrm{\Pi},
\end{equation}
where $\mathrm{\Pi}$ is hermitian. One may argue that \eref{eq:QL_proj_commutator} can not describe the general case, since one may use the
commutator not proportional to $\hbar$. But according to algorithm
complexity theory the choice of language consisting of the complete
set of operations plays no significant role: it can not change the
complexity class of the algorithm; see \eref{sec:eff} for details.
Using
\[
\forall\,k>0\quad \mathrm{P}_{\lambda}^k =
\mathrm{P}_{\lambda},\qquad \mathrm{P}_{\mu}^k = \mathrm{P}_{\mu},
\]
we obtain $\forall\,n>0$
\begin{eqnarray*}
\left(\mathrm{P}_{\lambda}\mathrm{P}_{\mu}\right)^n&=&
\left(\mathrm{P}_{\lambda}\mathrm{P}_{\mu}\right)^{n-1}\left(\mathrm{P}_{\mu}\mathrm{P}_{\lambda}
+i\hbar\mathrm{\Pi}\right)=\left(\mathrm{P}_{\lambda}\mathrm{P}_{\mu}\right)^{n-1}\left(\mathrm{P}_{\lambda}
+i\hbar\mathrm{\Pi}\right)\nonumber\\
&=&\cdots=\mathrm{P}_{\lambda}
\mathrm{P}_{\mu}\left(\mathrm{P}_{\lambda}+i\hbar\mathrm{\Pi}\right)^{n-1},
\end{eqnarray*}
where $n$ is integer. From the following
\begin{equation*}
\forall\,k\geq 0\quad\cases{
							\mathrm{P}_{\lambda}\left(i\hbar\mathrm{\Pi}\right)^{2k} =
								\left(i\hbar\mathrm{\Pi}\right)^{2k}\mathrm{P}_{\lambda}&\\
							\mathrm{P}_{\lambda}\left(i\hbar\mathrm{\Pi}\right)^{2k+1} =
								\left(i\hbar\mathrm{\Pi}\right)^{2k+1}\left(\mathrm{I}-\mathrm{P}_{\lambda}\right)&\\
						}
\end{equation*}
we obtain then
\begin{eqnarray*}
\fl\forall\,k\geq0\qquad\left(\mathrm{P}_{\lambda}+i\hbar\mathrm{\Pi}\right)^{2k}
&=&
	\left[\mathrm{P}_{\lambda}+(i\hbar\mathrm{\Pi})^2+i\hbar\mathrm{\Pi}\right]^k\\
&=&
	\sum_{s=0}^k\frac{k!}{(k-s)!s!}\left[\mathrm{P}_{\lambda}+(i\hbar\mathrm{\Pi})^2\right]^s(i\hbar\mathrm{\Pi})^{k-s}\\
&=&
	\sum_{s=0}^k\frac{k!}{(k-s)!s!}\Biggl[\mathrm{P}_{\lambda}\sum_{l=0}^s\frac{s!}{(s-l)!l!}(i\hbar\mathrm{\Pi})^{2(s-l)}\\
&&
	+\left(\mathrm{I}-\mathrm{P}_{\lambda}\right)(i\hbar\mathrm{\Pi})^{2s}\Biggr](i\hbar\mathrm{\Pi})^{k-s}\\
&=&
	\mathrm{P}_{\lambda}\left[\mathrm{I}+(i\hbar\mathrm{\Pi})^2+i\hbar\mathrm{\Pi}\right]^k
	+\left(\mathrm{I}-\mathrm{P}_{\lambda}\right)(i\hbar\mathrm{\Pi})^k\left(\mathrm{I}+i\hbar\mathrm{\Pi}\right)^k\\
&=&
	\mathrm{P}_{\lambda}\left(\mathrm{I}+\alpha\right)^k+\left(\mathrm{I}-\mathrm{P}_{\lambda}\right)\alpha^k,
\end{eqnarray*}
where
$\alpha=i\hbar\mathrm{\Pi}\left(\mathrm{I}+i\hbar\mathrm{\Pi}\right)$,
and finally results in
\begin{eqnarray}\label{eq:QL_nc_proj_n-power}
\fl\forall\,n>0\qquad\left(\mathrm{P}_{\lambda}\mathrm{P}_{\mu}\right)^n
=\cases{
		\beta\left(\mathrm{I}+\alpha\right)^k+\mathrm{P}_{\lambda}i\hbar\mathrm{\Pi}\alpha^k& for $ n=2k+1 $\\
		\beta\Bigl[\left(\mathrm{I}+\alpha\right)^k+i\hbar\mathrm{\Pi}\alpha^k\Bigr]+\gamma_{\,k}& for $ n=2(k+1) $\\
	},
\end{eqnarray}
where
$\beta=\mathrm{P}_{\mu}\mathrm{P}_{\lambda}+\left(\mathrm{I}-\mathrm{P}_{\lambda}\right)i\hbar\mathrm{\Pi}$
and
$\gamma_{\,k}=\mathrm{P}_{\lambda}\left(i\hbar\mathrm{\Pi}\right)^2\left(\mathrm{I}+\alpha\right)^k$.
It gives
\begin{equation}\label{eq:QL_n-proj_nc_limit}
\forall\,n>0\quad\lim_{\hbar\to0}\left(\mathrm{P}_{\lambda}\mathrm{P}_{\mu}\right)^n
= \lim_{\hbar\to0}\mathrm{P}_{\mu}\mathrm{P}_{\lambda}.
\end{equation}
Using \eref{eq:QL_proj_deq} and \eref{eq:QL_n-proj_nc_limit} one gets
\begin{eqnarray}\label{eq:QL_nc-conj_deq}
\lim_{\hbar\to0}\frac{\hbar}{i}\ln\left(\mathrm{P}_{\lambda}\wedge
\mathrm{P}_{\mu}\right)|\zeta\rangle
&=&
   \lim_{\hbar\to0}\frac{\hbar}{i}\ln\left(\mathrm{P}_{\mu}\mathrm{P}_{\lambda}\right)|\zeta\rangle
   =S_{\mu}+S_{{\mu}\to{\lambda}}+S_{{\lambda}\to\zeta}\nonumber\\
&=&S_{\lambda}+S_{{\lambda}\to{\mu}}+S_{{\mu}\to\zeta},
\end{eqnarray}
for which the following variations are true:
\begin{equation*}
\delta S_{\mu}=\delta S_{{\mu}\to{\lambda}}=\delta
S_{{\lambda}\to\zeta}=0,\qquad \delta S_{\lambda}=\delta
S_{{\lambda}\to{\mu}}=\delta S_{{\mu}\to\zeta}=0.
\end{equation*}
Expression \eref{eq:QL_nc-conj_deq} determines conjunction
dequantization for the non-commuting projectors.

Implication \eref{eq:QL_impl} by virtue of the previous result also
transforms into the classical one \eref{eq:CL_impl}:
\begin{equation}\label{eq:QL_impl_deq}
\mathrm{P}_{\lambda}\leq
\mathrm{P}_{\mu}\overset{\hbar\to0}{\dashrightarrow}\chi_{\lambda}\leq\chi_{\mu}.
\end{equation}

\section{Dequantization Information Loss Estimation}
\label{sec:eff}
Suppose that $S$ is in the pure quantum state $|\zeta\rangle$. Von
Neumann entropy $H_{\rm N}$ of the state
\begin{equation}\label{eq:QL_entr}
H_{\rm N}\left(|\zeta\rangle\right) = -\mathrm{Tr}\rho\ln\rho = 0.
\end{equation}
Here $\rho=|\zeta\rangle\langle\zeta|$ is the density matrix of the
system.

After dequantization the system $S$ can be described by the
corresponding characteristic function $\chi_{\lambda}$, see
\eref{eq:QL_proj_deq}, that splits the phase space $\Gamma_S$ into two
domains. As a result, $S$ can be characterized with Shannon entropy
$H_{\rm Sh}$
\begin{equation}\label{eq:CL_entr}
H_{\rm Sh}\left(\chi_{\lambda}\right)=
-\phi_{\lambda}\ln\phi_{\lambda}-\left(1-\phi_{\lambda}\right)\ln\left(
1-\phi_{\lambda}\right),\qquad \phi_{\lambda}=
\frac{\int\mathcal{D}x\chi_{\lambda}}{\int\mathcal{D}x}.
\end{equation}
In the following the argument of $H_{\rm Sh}$ may be denoted with the
characteristic function or the corresponding projectors with no
change in the expression meaning.

Thus after dequantization entropy depends on how $\chi_{\lambda}$
splits $\Gamma_S$. It is nonzero except when $\phi_{\lambda}=0$ or
$\phi_{\lambda}=1$. One may notice that the entropy is upper bounded,
i.e.
\begin{equation}\label{eq:CL_entr_up-bnd}
\forall\,\lambda\quad H_{\rm Sh}\left(\chi_{\lambda}\right)\leq\ln2.
\end{equation}
The existence of the upper bound means that some quantum states
after the dequantization lose all quantum correlations causing the
maximal information loss possible.

Any logic statement consisting of commuting projectors is equivalent
to some projector. Consequently, any pure quantum state under the
statement transforms to another pure state leaving the von Neumann
entropy $H_{\rm N}$ unchanged. However, after statement dequantization the
entropy will change because of re-splitting $\Gamma_S$. To show
this, the entropy of dequantized logic operations should be
explored.

Conjunction entropy of \emph{commuting} projectors after taking the
limit $\hbar\to0$, see \eref{eq:QL_c-conj_deq}, is defined as
\begin{equation}\label{eq:QL_c-conj_deq_entr}
H_{\rm Sh}\left(\chi_\wedge\right)=-\phi_\wedge\ln\phi_\wedge-\left(1-\phi_\wedge\right)\ln\left(1-\phi_\wedge\right)
\leq\ln2
,\qquad \phi_\wedge = \frac{\int\mathcal{D}x\chi_\wedge}{\int \mathcal{D}x},
\end{equation}
where $\chi_\wedge = \chi_{\lambda}\chi_{\mu}$. Since $\chi_\wedge$ is nothing more but some characteristic function, we used expression \eref{eq:CL_entr_up-bnd} to define the upper bound on $H_{\rm Sh}\left(\chi_\wedge\right)$.

For the quantum negation entropy after dequantization \eref{eq:QL_neg_deq} we obtain
\begin{equation}\label{eq:QL_neg_deq_entr}
H_{\rm Sh}\left(\chi_{\neg\lambda}\right) = H_{\rm Sh}\left(1 -
\chi_{\lambda}\right) = H_{\rm Sh}\left(\chi_{\lambda}\right).
\end{equation}
Expression \eref{eq:QL_neg_deq_entr} means that because of the
symmetry of \eref{eq:CL_entr} negation does not change entropy in
statement even after the dequantization.

For the implication of commuting projectors one gets that, according
to \eref{eq:CL_impl} and \eref{eq:QL_impl}, after the logic conversion
\eref{eq:QL_impl_deq} entropy will have the following property:
\begin{equation}\label{eq:QL_c-impl_deq_entr}
\mathrm{P}_{\lambda}\leq \mathrm{P}_{\mu}\qquad\Rightarrow\qquad
H_{\rm Sh}\left(\chi_\wedge\right)=H_{\rm Sh}\left(\chi_{\lambda}\right),
\end{equation}
where $\chi_\wedge=\chi_{\lambda}\chi_{\mu}$.

As before, in case of non-commuting projectors it is enough to
consider entropy of the corresponding conjunction \eref{eq:QL_nc-conj}
after the logic conversion \eref{eq:QL_nc-conj_deq}.

Let $P_{\lambda},P_{\mu}$ be two non-commuting projectors satisfying
\eref{eq:QL_proj_commutator}. The initial state $|\zeta\rangle$ of the
system can be expanded into series by the eigenstates of commutator
$ \mathrm{\Pi} $:
\[
|\zeta\rangle =
\sum_\pi^{\mathrm{dim}\mathrm{\Pi}}\zeta_\pi|\pi\rangle,
\qquad\mathrm{\Pi}|\pi\rangle = \pi|\pi\rangle.
\]
Terms containing nonzero powers of $\mathrm{\Pi}$ will vanish in
accordance with \eref{eq:QL_nc_proj_n-power} while taking the limit
$\hbar\to0$ in conjunction \eref{eq:QL_nc-conj}. Thus density matrix
$\rho$ should be traced over the eigenstates of $\mathrm{\Pi}$.
Under this averaging pure state transforms into the mixture for
which the von Neumann entropy is nonzero:
\begin{eqnarray}\label{eq:QL_nc-conj_entr_Pi-tr}
H_{\rm N}\left(|\zeta\rangle\right)\to
H_{\rm N}\left(\rho_\mathrm{\Pi}\right)=-\mathrm{Tr}\rho_\mathrm{\Pi}\ln\rho_\mathrm{\Pi}
=-\sum_\pi^{\mathrm{dim}\mathrm{\Pi}}|\zeta_\pi|^2\ln|\zeta_\pi|^2\leq\ln\mathrm{dim}\mathrm{\Pi}.
\end{eqnarray}
Here
$\rho_\mathrm{\Pi}=\mathrm{Tr}_\mathrm{\Pi}|\zeta\rangle\langle\zeta|$.

In addition, the contribution of every eigenstate $|\pi\rangle$ from
the mixture $\rho_\mathrm{\Pi}$ to the whole entropy should be
included. Any such term is expressed similarly to \eref{eq:CL_entr}
\begin{equation}
H_{\rm Sh}\left(\chi_{\wedge_\mathrm{\Pi}|\pi}\right) =
-\phi_{\wedge|\pi}\ln\phi_{\wedge|\pi} -
\left(1-\phi_{\wedge|\pi}\right)\ln\left(1-\phi_{\wedge|\pi}\right),
\end{equation}
where $\chi_{\wedge_\mathrm{\Pi}}$ is the characteristic function
corresponding to the conjunction of our projectors and
\begin{equation*}
	\phi_{\wedge|\pi} = \frac{\int\mathcal{D}x_{|\pi}\chi_{\lambda}\chi_{\mu}}{\int\mathcal{D}x_{|\pi}}.
\end{equation*}
Here and in the following subscript $_{|\pi}$ means that the transition starting from
the state $|\lambda\rangle$ or $|\mu\rangle$ results in the
corresponding state $|\pi\rangle$ but not in $|\zeta\rangle$ as
before, see \eref{eq:QL_proj}.

Summarizing, the whole entropy for the dequantized conjunction of
two non-commuting projectors is
\begin{equation}\label{eq:QL_nc-conj_deq_entr}
H\left(\chi_{\wedge_\mathrm{\Pi}}\right) =
H_{\rm N}\left(\rho_\mathrm{\Pi}\right) +
\sum_\pi^{\mathrm{dim}\mathrm{\Pi}}|\zeta_\pi|^2H_{\rm Sh}\left(\chi_{\wedge_\mathrm{\Pi}|\pi}\right).
\end{equation}
There is no subscript $ _{\rm Sh} $ nor $ _{\rm N} $ on the lhs of \eref{eq:QL_nc-conj_deq_entr} since it is a sum of both the von Neumann and Shannon entropies.
As one can see, \eref{eq:QL_c-conj_deq_entr} is easily obtained via
formal setting $\mathrm{dim}\mathrm{\Pi}=1$ in \eref{eq:QL_nc-conj_deq_entr}. The upper bound of
$H\left(\chi_{\wedge_\mathrm{\Pi}}\right)$, see \eref{eq:CL_entr_up-bnd} and \eref{eq:QL_nc-conj_entr_Pi-tr}, is
\begin{equation}\label{eq:QL_nc-conj_deq_entr_up-bnd}
H\left(\chi_{\wedge_\mathrm{\Pi}}\right)\leq\ln\mathrm{dim}\mathrm{\Pi}
+ \ln2.
\end{equation}

Now we can consider the case of the commutators not proportional to
$\hbar$ in details, see \eref{eq:QL_proj_commutator} and the text
right after it. To do it one can replace $\hbar{\rm \Pi}$ in \eref{eq:QL_proj_commutator} by some hermitian operator $C$. Such an operator
can be diagonalized, i.e. represented in the form
$C=\sum_1^{\mathrm{dim}C} cP_c$, where $P_c$ is the projector on the
eigenstate of $C$ with eigenvalue $c$. Now, following the
dequantization procedure for such an operator (see \eref{sec:deq}) one will result in re-definition of coefficients
$\alpha,\beta,\gamma$ in \eref{eq:QL_nc_proj_n-power} without any
change in \eref{eq:QL_n-proj_nc_limit} and in \eref{eq:QL_nc-conj_deq_entr}. In other words, one again will meet with the information loss
while taking the semiclassical limit $\hbar\to0$ without any change
at the end and hence with no additional entropy except the estimated
one.

Implication of the non-commuting operators is similar to analysis of the commuting ones, see \eref{eq:QL_c-impl_deq_entr}. The only difference is
that the non-commuting conjunction entropy \eref{eq:QL_nc-conj_deq_entr} should be used, i.e.
\begin{equation}\label{eq:QL_nc-impl_deq_entr}
\mathrm{P}_{\lambda}\leq\mathrm{P}_{\mu}\qquad\Rightarrow\qquad
H\left(\chi_{\wedge_\mathrm{\Pi}}\right)=H_{\rm Sh}\left(\chi_{\lambda}\right),
\end{equation}
where projectors satisfy \eref{eq:QL_proj_commutator}. However, this
is the generalization of \eref{eq:QL_c-impl_deq_entr}; the latter is
obtained by setting $\mathrm{dim}\mathrm{\Pi}=1$ in \eref{eq:QL_nc-impl_deq_entr} as we did it before.

The obtained results define the entropy increase for any elementary
logical statements under the logic conversion. Such elementary
statements are atomic, and thus are equivalent to the one-qubit
register. But for the complete analysis of the information gap
registers of arbitrary length should be observed.

Let $|\zeta\rangle^{\otimes N_{\mathrm{I}}}$ be an
$N_{\mathrm{I}}$-qubit register. Any calculation with it is
equivalent to construction of some logical expression
$\mathbb{E}_{\mathrm{I}}$ from the elementary logical operations
defined on projectors. Suppose that $\mathbb{E}_{\mathrm{I}}$ has no
implications inside (that's underlined with index $_{\mathrm{I}}$)
and consists of $n_{\mathrm{I}}$ negations $\neg$ and
$c_{\mathrm{I}}$ conjunctions $\wedge$. The following expression
\[
N_{\mathrm{I}}\leq n_{\mathrm{I}}+c_{\mathrm{I}}
\]
must be true since else such a coding can be applied where
$N_{\mathrm{I}}-n_{\mathrm{I}}-c_{\mathrm{I}}$ qubits will be
obsolete.

Conjunctions $c_{\mathrm{I}}$ are defined on the non-commuting
projectors in general. Thus one has to include all commutator
\eref{eq:QL_proj_commutator} contributions while estimating the
entropy. After neglecting the first such commutator all subsequent
elementary statements will operate on the mixture but not on the
pure state. However, as negation does not influence the entropy, the
conjunctions operating on the mixture should be observed only.

Suppose that the expression
$\mathbb{E}_{\mathrm{I},\mathrm{\Pi}_{2}\mathrm{\Pi}_{1}}$ consists
of two conjunctions characterized with commutators
$\mathrm{\Pi}_{1}$ (corresponds to the first calculated conjunction)
and $\mathrm{\Pi}_{2}$ (the second one). After dequantization
entropy of the expression will be
\[
H\left(\mathbb{E}_{\mathrm{I},\Pi_{2}\Pi_{1}}|\zeta\rangle\right)
= H\left(\chi_{\wedge_{\mathrm{\Pi}_{1}}}\right) +
\sum_{\pi_{1}}^{\mathrm{dim}\mathrm{\Pi}_{1}}|\zeta_{\pi_{1}}|^2H_{\rm Sh}\left(\chi_{\wedge_{\mathrm{\Pi}_{2}}|\pi_{1}}\right).
\]
In general, for $\mathbb{E}_{\mathrm{I}}$ on the register
$|\zeta\rangle^{\otimes N_{\mathrm{I}}}$ the whole entropy  will be
estimated by recurrent formula
\begin{eqnarray}\label{eq:QL_N-reg_n+c_deq_entr}
H\left(\mathbb{E}_{\mathrm{I}}|\zeta\rangle^{\otimes
N_{\mathrm{I}}}\right) =
\sum_{i=1}^{q_{\mathrm{I}}}H_{\rm Sh}\left(\chi_{\lambda_i}\right)+H\left(\chi_{\wedge_{\mathrm{\Pi}_{1}}}\right)
+\sum_{\pi_1}^{\mathrm{dim}\mathrm{\Pi}_1}|\zeta_{\pi_1}|^2H_{\rm Sh}\left(\chi_{\wedge_{\mathrm{\Pi}_{2}}|\pi_1}\right).
\end{eqnarray}
Here $q_{\mathrm{I}}$ is the number of qubits equipped in no
conjunction. Using \eref{eq:CL_entr_up-bnd} and \eref{eq:QL_nc-conj_deq_entr_up-bnd}, one may obtain the upper bound for the entropy:
\begin{equation}\label{eq:QL_N-reg_n+c_deq_entr_up-bnd}
H\left(\mathbb{E}_{\mathrm{I}}|\zeta\rangle^{\otimes
N_{\mathrm{I}}}\right)\leq\left(q_{\mathrm{I}}+c_{\mathrm{I}}\right)\ln2
+ \sum_{k=1}^{c_{\mathrm{I}}}\ln\mathrm{dim}\mathrm{\Pi}_k.
\end{equation}

To estimate the entropy of some general expression $\mathbb{E}$ one must count
over all implications made during the calculation. It means that for
$\mathbb{E}$ containing subexpressions
$\{\mathbb{E}_{\mathrm{I}}\}_\mathrm{I}$ on the register
$|\zeta\rangle^{\otimes N}$ total entropy
$H\left(\mathbb{E}|\zeta\rangle^{\otimes N}\right)$ must consist
of contributions from all the subexpressions, each of which is defined by \eref{eq:QL_N-reg_n+c_deq_entr_up-bnd}.

\section{Conjunction Theorem}
\label{sec:thm}
Now we are almost ready to verify our approach on real algorithms.
However, any quantum algorithm, to be the computable one in finite
time, should not contain the conjunction of non-commuting projectors
since it requires an infinite time for its construction, see
\eref{eq:QL_nc-conj}. The algorithm should use the finite products of
non-commuting projectors instead. So there is a need in setting an
interrelation between the non-commuting conjunction and the finite
product of the non-commuting projectors. We provide this with the
help of the following theorem.

\textbf{Theorem.}\label{thm} 
	Let $\mathrm{P}_{\lambda},
	\mathrm{P}_{\mu}$ be any two projective operators such that
	$\left[\mathrm{P}_{\lambda},\mathrm{P}_{\mu}\right]=i\hbar\mathrm{\Pi}$,
	$\mathrm{P}_{\wedge}=\mathrm{P}_{\lambda}\wedge\mathrm{P}_{\mu}$.
	Then
	\begin{equation}\label{eq:th_1}
		\forall\,k>0,\quad
		H\bigl(\left(\mathrm{P}_{\lambda}\mathrm{P}_{\mu}\right)^k\bigr)=H\left(\mathrm{P}_{\wedge}\right).
	\end{equation}

\textbf{Proof.}
 From \eref{eq:QL_nc-conj} we can write
 \begin{equation}\label{eq:th_2}
  \left(\mathrm{P}_{\lambda}\mathrm{P}_{\mu}\right)^k\mathrm{P}_{\wedge}=
  \mathrm{P}_{\wedge}\left(\mathrm{P}_{\lambda}\mathrm{P}_{\mu}\right)^k=\mathrm{P}_{\wedge}.
 \end{equation}
 As it follows from \eref{eq:QL_nc-conj},
 $\mathrm{P}_{\wedge}^2=\mathrm{P}_{\wedge}$, i.e. it is a
 projective operator. This is not true for the product
 $\left(\mathrm{P}_{\lambda}\mathrm{P}_{\mu}\right)^k$, since
 $\mathrm{P}_{\lambda}$ and $\mathrm{P}_{\mu}$ do not commute. But,
 such the product defines some Hilbert subspace, and therefore
 \eref{eq:th_2} is the common implication for commuting operators, see
 \eref{eq:QL_impl}. Then, using \eref{eq:QL_c-impl_deq_entr} finally we
 result in \eref{eq:th_1} completing the proof. \opensquare

Now we can generalize expressions \eref{eq:QL_N-reg_n+c_deq_entr}
and \eref{eq:QL_N-reg_n+c_deq_entr_up-bnd}. Due to \eref{thm} any conjunction gives the same IL as the product of the
projectors involved in it. In such a case, the implication
(commuting or non-commuting) IL can be estimated directly: due to
definition it contains the conjunction or the projector replacing
the conjunction itself. All we need to do is just to remove the
subscript $_\mathrm{I}$ in \eref{eq:QL_N-reg_n+c_deq_entr} and
\eref{eq:QL_N-reg_n+c_deq_entr_up-bnd}. So, finally we obtain
\begin{eqnarray}\label{eq:QL_IL_final}
H\left(\mathbb{E}|\zeta\rangle^{\otimes N}\right) =
\sum_{i=1}^{q}H_{\rm Sh}\left(\chi_{\lambda_i}\right)+H\left(\chi_{\wedge_{\mathrm{\Pi}_{1}}}\right)
+\sum_{\pi_1}^{\mathrm{dim}\mathrm{\Pi}_1}|\zeta_{\pi_1}|^2H_{\rm Sh}\left(\chi_{\wedge_{\mathrm{\Pi}_{2}}|\pi_1}\right)
\end{eqnarray}
and
\begin{equation}\label{eq:QL_IL_final_up-bnd}
H\left(\mathbb{E}|\zeta\rangle^{\otimes
N}\right)\leq\left(q+c\right)\ln2 +
\sum_{k=1}^{c}\ln\mathrm{dim}\mathrm{\Pi}_k,
\end{equation}
where $\mathbb{E}$ is the expression being processed on the
$N$-qubit register, $q$ is the number of qubits equipped in no
conjunction, and $c$ is the number of conjunctions; any product of
non-commuting operators should be considered as conjunction due to
\eref{thm}.

\section{Examples}
\label{sec:ex}
Let us introduce some notations before we proceed. At first we
define the following projectors:
\begin{equation*}
\mathrm{P}_q=|q\rangle\langle q|,\qquad\mathrm{P}_{\neg
q}=\mathrm{I}-\mathrm{P}_q,\qquad|q\rangle\colon\sigma_q|q\rangle=|q\rangle,
\end{equation*}
where $\mathrm{I}$ is the unit operator, $q=\{x,y,z\}$ and
$\sigma_q$ is the corresponding Pauli matrix. It is easy to check
that $\mathrm{P}_q\mathrm{P}_{q'}\neq\mathrm{P}_{q'}\mathrm{P}_q$ if
$q\neq q'$.

Since $\{\mathrm{I},P_x,P_y,P_z\}$ are linearly independent, we can
encode any qubit operator as some linear combination of these
matrices. To proceed we need the following operators:
\begin{eqnarray*}
&&\mathrm{W}_k=\frac{1}{\sqrt{2}}\left(\mathrm{P}_{zk}-\mathrm{P}_{\neg
zk}+\mathrm{P}_{xk}-\mathrm{P}_{\neg
xk}\right)=\sqrt{2}\left(\mathrm{P}_{xk}-\mathrm{P}_{\neg zk}\right),\\
&&\mathrm{C}_{k,s}=(1-e^{i\phi_{k,s}})\left(\mathrm{P}_{zs}\mathrm{P}_{\neg
zk}+\mathrm{I}_s\mathrm{P}_{zk}\right)+e^{i\phi_{k,s}}\mathrm{I}_s\mathrm{I}_k,
\end{eqnarray*}
where $\mathrm{W}_k$ is the Walsh-Hadamard gate on the $k$-th qubit,
and $\mathrm{C}_{k,s}$ is the controlled-phase gate on the $k$-th
and $s$-th qubits with phase shift $\phi_{k,s}=\pi/2^{s-k}$. Here
and in the following the operator's subscripts $_{k,s}$ denote the
qubits these operators act on.

We emphasize that using another basis matrices (and consequently the
projectors) will not influence the result, since may be provided by
simple unitary rotation of the one presented. This is simple
consequence of the fact that changing the language (but except the
unary languages consisting of one symbol only) can not significantly
influence the algorithm complexity.

Now we can use \eref{eq:QL_IL_final_up-bnd} for estimation of the IL
of any quantum algorithm $\mathbb{E}$. Processing the estimation we
should keep in mind that the number of conjunctions $c$ equals to
the number of projector's intersections from the viewpoint of the IL
(see \eref{thm}).

Below we estimate IL for two different quantum algorithms: quantum
discrete Fourier transform FFT$_\mathrm{Q}$ and Grover search
algorithm Gr$_\mathrm{Q}$. Those who are interested in the details
of the algorithms we refer to \cite{qc_nielsen}. These
algorithms provide essential speed-up compared to their classical
analogs, which are exponentially complex, and one expects that to be
reflected by IL in some way.

\subsection{FFT$_\mathrm{Q}$ dequantization}
\label{sec:ex-fft}
As it is known, FFT$_\mathrm{Q}$ on the $N$-qubit register may be
written as the following operator:
\begin{equation}\label{FFTq}
\mathrm{FFT}_\mathrm{Q}=\mathrm{\Phi}_0\cdots\mathrm{\Phi}_{N-1},\qquad
\mathrm{\Phi}_k=\mathrm{W}_k\mathrm{C}_{k,N-1}\mathrm{C}_{k,N-2}\cdots
\mathrm{C}_{k,k+1}.
\end{equation}
One can notice that every $\mathrm{C}_{k,s}$ contains 2 non-reducing
terms with $\mathrm{P}_{zk}$ which do no commute with the
corresponding $\mathrm{P}_{xk}$ of $\mathrm{W}_k$. Then the number
of non-commuting projector products is $c_k=2^{N-k-1}$ for any
$\mathrm{\Phi}_k$ and
\begin{equation}\label{eq:FFTq_ci_est_2}
c=\sum_{k=0}^{N-1}c_k=\sum_{k=0}^{N-1}2^{N-k-1}=2^N-1.
\end{equation}
Any $\mathrm{\Phi}_k$ contains $N-k-1$ commuting projector products
(commuting conjunctions); summing over $k$ gives $N(N-1)/2$
commuting conjunctions in general. Substituting this, \eref{eq:FFTq_ci_est_2} and $\mathrm{dim}\mathrm{\Pi}_k=2^N$ in \eref{eq:QL_IL_final_up-bnd}, we obtain then
\begin{equation}\label{eq:FFTq_deq}
H\left(\mathrm{FFT}_{\mathrm{Q}}\right)\leq\Bigl[q+\frac{N(N-1)}{2}+(N+1)\left(2^N-1\right)\Bigr]\ln2=
\mathcal{O}\left(N2^N\right),
\end{equation}
thus meeting an exponential IL of the dequantized FFT$_\mathrm{Q}$.
As it is known, its classical analog FFT$_\mathrm{C}$ needs
$\mathcal{O}\left(N2^N\right)$ amount of resources.

\subsection{Grover dequantization}
\label{sec:ex-gr}
$\mathrm{Gr}_\mathrm{Q}$, which is operating on the database
containing $2^N$ elements, can be represented with the following
operator
\begin{eqnarray}\label{Grq}
\mathrm{Gr}_\mathrm{Q}
&=&
   \Bigl\{\left[2\left(\mathrm{W}\mathrm{P}_z\mathrm{W}\right)^{\otimes
   N}-\mathrm{I}^{\otimes N}\right]\otimes\left(\mathrm{P}_{\neg
   z}-\mathrm{P}_z\right)\Bigr\}^{\frac{\pi}{4}2^{N/2}}\mathrm{U}_\Gamma,\\
\mathrm{U}_\Gamma
&\colon&
   |x\rangle|0\rangle\to|x\rangle|\Gamma(x)\rangle,\nonumber
\end{eqnarray}
where $\Gamma$ is the tested statement (i.e.
$\mathrm{Gr}_\mathrm{Q}$ determines the elements on which $\Gamma$
is true). The operator $\mathrm{U}_\Gamma$ requires a number of
gates depending on particular expression for $\Gamma$, and thus will
not be considered in the following.

As for the component $\left(\mathrm{P}_{\neg z}-\mathrm{P}_z\right)$
acting on the ancillary qubit, it includes $c_{k|\Gamma}=1$
intersections for the complementary (and hence commuting) projectors
only, for which one can put formally
$\mathrm{dim}\mathrm{\Pi}_{k|\Gamma}=1$ while estimating IL. The
number of these ancillary intersections is
\begin{equation}\label{eq:Grq_ci_Gamma_est}
c_{|\Gamma}=\sum_{k=1}^{\frac{\pi}{4}2^{N/2}}c_{k|\Gamma}=\frac{\pi}{4}2^{N/2}.
\end{equation}
For operator in the square brackets we obtain that
\[
\mathrm{W}\mathrm{P}_z\mathrm{W}=2\left(\mathrm{P}_x-\mathrm{P}_{\neg
z}\right)\mathrm{P}_z\left(\mathrm{P}_x-\mathrm{P}_{\neg
z}\right)=2\mathrm{P}_x\mathrm{P}_z\mathrm{P}_x,
\]
thus giving one intersection of non-commuting projectors. The number
of such intersections in the square brackets is $c_{k|[\,]}=N$ (one
for every qubit in the register). Since iteration should be applied
$\frac{\pi}{4}2^{N/2}$ times, then
\begin{equation}\label{eq:Grq_ci_D_est}
c_{|[\,]}=\sum_{k=1}^{\frac{\pi}{4}2^{N/2}}c_{k|[\,]}=\sum_{k=1}^{\frac{\pi}{4}2^{N/2}}N=\frac{\pi}{4}N2^{N/2}.
\end{equation}
As $\mathrm{dim}\mathrm{\Pi}_{k|[\,]}=2^N$, we obtain after
substituting \eref{eq:Grq_ci_Gamma_est} and \eref{eq:Grq_ci_D_est} into
\eref{eq:QL_IL_final_up-bnd}
\begin{eqnarray}\label{eq:Grq_def}
H\left(\mathrm{Gr}_\mathrm{Q}\right)
&\leq&
   \left(q+c_{|\Gamma}\right)\ln2+c_{|[\,]}\left(1+N\right)\ln2\nonumber\\
&=&
   \left[q+\frac{\pi}{4}\left(N^2+N+1\right)2^{N/2}\right]\ln2=\mathcal{O}\left(N^22^{N/2}\right).
\end{eqnarray}
As it is known, classical search algorithm requires
$\mathcal{O}\left(2^N\right)$ number of resources, while
$\mathrm{Gr}_\mathrm{Q}$ needs $\mathcal{O}\left(2^{N/2}\right)$.
Thus, as we see, we obtain non-polynomial IL in this case. It may be
an example of the ``not complete'' algorithm reduction, i.e. when
the algorithm under the dequantization reduces to the rather
complicated one.

As the search algorithm belongs to the NP complexity class, the
example demonstrates that at least some quantum algorithms being NP
(here this is $\mathrm{Gr}_\mathrm{Q}$) do not meet complete IL
(i.e. IL for them does not necessarily equal in the number of
resources required with their classical analogs) under
dequantization: here we obtained
$\mathcal{O}\left(N^22^{N/2}\right)$ instead of
$\mathcal{O}\left(2^N\right)$ IL. The reason for such a difference
is the transformation of the algorithm which is discussed in the
next section.

One may argue that the obtained results may be explained by the Gottesman-Knill theorem (see \cite{qc_nielsen} for the formulation and proof): \eref{FFTq} includes the gates outside the Clifford group, while \eref{Grq} employs the gates from the Clifford group only. And this is what we obtained by expressions \eref{eq:FFTq_deq} and \eref{eq:Grq_def}. However, compared to the Gottesman-Knill theorem, we exactly show how the computational efficiency is being reduced and give concrete recipe for the loss estimation, see \eref{eq:QL_IL_final} and \eref{eq:QL_IL_final_up-bnd}.

\section{Discussion}
\label{sec:dis}
In the manuscript we dequantized the complete set of elementary
quantum logic operations including the non-commuting conjunction. We
calculated the IL and its upper bound for the operations after
taking the semiclassical limit. We derived the general expression
estimating the IL for any dequantized quantum algorithm, see
\eref{eq:QL_IL_final} and \eref{eq:QL_IL_final_up-bnd}. We formulated
and proved theorem (see Section \ref{sec:thm}), which is necessary for estimation of conjunction of
non-commuting projectors. Finally, the obtained results were applied
for IL estimation for FFT$_\mathrm{Q}$ and Gr$_\mathrm{Q}$
algorithms. The developed technique demonstrated exponential
(\eref{sec:ex-fft}) and non-polynomial (\eref{sec:ex-gr}) IL for the
algorithms correspondingly.

Expression \eref{eq:QL_IL_final} estimates amount of information
being lost by quantum algorithm, which is encoded with $\mathbb{E}$,
after processing through the semiclassical limit. It implies that at
least the \emph{description} of the IL requires the additional
memory of the $H\left(\mathbb{E}|\zeta\rangle^{\otimes N}\right)$
size which is upper bounded with \eref{eq:QL_IL_final_up-bnd}. At the
same time, such a description requires at least the same increase of
amount of elementary logical steps (by one per each additional
memory cell to write it down). So we conclude that the technique
presented in the paper might shed some light on the NP problem (in
case we consider some NP-complete algorithm, see \eref{sec:ex-gr}) and
on the algorithm complexity classification.

Any quantum algorithm under dequantization keeps the number of
elementary logical operations the same with no change in efficiency
in its common sense. But the description of its IL requires
additional memory and, consequently, time (measured in the number of
elementary logical steps). The interrelation between both the
efficiencies is unclear, since the dequantized algorithm and its IL
description do not coincide.

The algorithmic entropies, see \cite{it_alg_e}, may be used
to describe ``distance'' between the desired and the calculated
result in case of using quantum or classical algorithm.  The
entropies are used to estimate the probability of obtaining the
desired result; they are defined for the states calculated with some
algorithm. Our approach differs a lot from this one since we
investigate the changes of the elementary logic operations while
taking the limit $\hbar\to0$.

The Kolmogorov complexity approach is useful for estimation of the
difference between quantum and classical calculation. It gives the
minimized in size program that realizes the corresponding algorithm.
Such an approach helps to define conditions on the calculations,
which are easy in quantum but are hard in classical case. For more
details see \cite{qc_k_i_def} and other ones \cite{qc_k_intro,qc_k_alg,qc_k_appl}.

However, our approach differs from the Kolmogorov's one.
Dequantization of elementary QL operations allows to estimate the
corresponding entropy for any logical expression. It gives the
amount of information loss during the \emph{reduction} of quantum
algorithm to the classical one. It has much in common with (but can
not be interpreted as comparison of) the corresponding Kolmogorov
complexities for the quantum and classical algorithms solving
\emph{the same} problem. The number of elementary operations does
not change when in the semiclassical limit. The similarity origins
from the re-estimation of quantum gates in terms of classical ones.
But, after the dequantization algorithm may solve another problem
(like FFT$_\mathrm{Q}$), thus pointing out the differences with the
Kolmogorov approach.

We illustrate this statement with the help of the classical discrete
fast Fourier transform (FFT). As it is known, FFT$_\mathrm{Q}$ is a
polynomial time algorithm. It needs $\mathcal{O}\left(N^2+N\right)$
operations, while FFT needs $\mathcal{O}\left(N2^N\right)$.
According to \eref{sec:ex}, the number of elementary
operations during dequantization remains the same, i.e. polynomial.
However, some amount of information is lost, and this amount can be
estimated. We suggest that the only explanation for this is the
algorithm changeover. In particular FFT$_\mathrm{Q}$ transforms into
the Legendre transform (but not into FFT) \cite{id_repr}; for some
more information see \cite{id_cplx}.

Such algorithm simplification after dequantization is explained by
the fact that QL algebra can be split up on some Boolean
subalgebras, each of which is similar to the CL
algebra \cite{ql_context}. But statements from different QL
subalgebras do not commute, thus providing the largest IL possible.

It has been widely believed that entanglement is a quantum resource
responsible for high efficiency of quantum algorithms. In our
approach the (non-commuting in general) projective operators are
used only, with no direct relation to entanglement. One may say that
the non-commuting projectors project the state to different
subspaces (say ${\rm H}_1$ and ${\rm H}_2$) such that the basis vectors from
${\rm H}_1$ are represented as entangled in the basis of $H_2$. But there
are doubts this can be stated and proved in general. In \cite{en_shor} a simulation of the Shor's factoring algorithm was
made, and the authors found no significant role of entanglement in
providing the exponential speed-up of the algorithm. Based on this
and on our own results, we suppose that entanglement can not be
considered as the resource of the computational speed-up in quantum
calculation, and state that high computational efficiency of quantum
algorithms is highly interrelated with the presence of non-commuting
statements which can not be simulated efficiently by CL. However, the last item is true if IL is strongly
interconnected with the computational efficiency only and requires further research. Both the efficiencies, in spite
of having much in common, differ from each other.

Some questions still remain open, and it seems reasonable to solve
them. Here they are:
\begin{itemize}
\item How are the efficiency and IL are
interconnected with each other?

\item If some optimal quantum algorithm gives an exponential
IL, whether it implies that the classical algorithm for the same
problem is exponential in time?

\item Can the presented approach be used for the comparison of
quantum and classical algorithm complexity classes, and correlation
ascertainment between these classes?

\item Generally any quantum algorithm transforms into another one,
that solves another problem, under the semiclassical limit. But,
what about the reversion: can one obtain some quantum algorithm (or
the class of them), being given the classical one? Some
investigation on the topic of transition from subsets of CL to
compatible (i.e. determined with mutually commuting operator sets)
linear subspaces of QL are presented in paper \cite{ql_deq_lift};
it is called lifting in the manuscript. In our opinion, such a
reversion should be ambiguous due to the differences between subsets
and linear subspaces. The point is that there is no recipe to go to
the incompatible subspaces. In our opinion, formalization and
further development of the approach presented in \cite{ql_relation} might be helpful while constructing the
recipe. However it is not known for sure whether it can be solved or
not at least for some classes of quantum algorithms. In particular
one can try to build the quantum analog of the Legendre transform.
Due to ambiguity one is expected to derive some class of quantum
algorithms but not FFT$_\mathrm{Q}$ only. One more interesting point
is to look for the quantum analogs of inefficient classical
algorithms such as FFT or factorization and to verify whether the
analogs will be inefficient in QL too.
\end{itemize}
Summing up the questions mentioned above we conclude that the
problem of the reverse transition to dequantization is worth of
further investigation.

\ack
	We are thankful to E. D. Belokolos for all the valuable discussions
	during preparation of this work and to O. Fialko for all the remarks
	which helped to improve it a lot.
	
\bibliographystyle{unsrt}
\bibliography{D:/Bibliography/Catalogue/bib}

\end{document}